\documentclass[12pt]{article}
\usepackage[utf8]{inputenc}
\usepackage{cite}
\usepackage{graphicx}
\usepackage{amsmath}
\usepackage[margin=0.9in]{geometry}
\usepackage{bm}
\usepackage{authblk}
\title{Exactly solvable 1D model explains the low-energy vibrational level structure of protonated methane}

\author[1]{Jonathan I. Rawlinson \thanks{jonathanianrawlinson@gmail.com}}
\author[2,3]{Csaba F\'abri}
\author[2,3]{Attila G. Cs\'asz\'ar}

\affil[1]{School of Mathematics, University of Bristol, Bristol, UK}
\affil[2]{Laboratory of Molecular Structure and Dynamics, 
Institute of Chemistry, E\"otv\"os Lor\'and University, P\'azm\'any P\'eter s\'et\'any 1/A, H-1117 Budapest, Hungary}
\affil[3]{MTA-ELTE Complex Chemical Systems Research Group, P.O. Box 32, H-1518 Budapest 112, Hungary}

\begin{document}

\maketitle

\begin{abstract}
    A new one-dimensional model is proposed for the low-energy vibrational 
quantum dynamics of CH$_{5}^{+}$\ based on the motion of an effective particle 
confined to a $60$-vertex graph $\Gamma_{60}$
with a single edge length parameter.
Within this model,
the quantum states of CH$_{5}^{+}$\ are obtained in analytic form and are related 
to combinatorial properties of $\Gamma_{60}$. 
%In particular, we show that 
The bipartite structure of $\Gamma_{60}$
gives a simple explanation for curious symmetries observed in 
numerically exact variational calculations on CH$_{5}^{+}$
\end{abstract}

Protonated methane, CH$_{5}^{+}$, also called methonium, 
is considered to be the prototype of pentacoordinated nonclassical carbonium 
ions \cite{68OlSc,69OlKlSc,74Olah}.
The curious carbonium cations yielded an extremely rich chemistry 
and a Nobel-prize to their discoverer, George Olah \cite{94Olah}.
Nevertheless, these are not the only sources of fame 
for carbonium ions and in particular for CH$_{5}^{+}$.
%In fact, o
Over the last  two decades \cite{99WhTaOk}, 
the internal motion of CH$_{5}^{+}$\  has been posing 
a formidable challenge to high-resolution spectroscopists 
\cite{99WhTaOk,05AsKuRe,08WaCa,10IvAsWiHu,15AsYaBrPo,15WoMa,15ScScJe,16WaCa,17BrScAs,17FaQuCs,18FaCs}.
% and resulted in many developments \cite{16WaCa} without a full understanding.
The most outstanding issue is that 
the observed spectra of CH$_{5}^{+}$\  remain exceptionally complex even when they are
observed at temperatures of a few K \cite{15AsYaBrPo,17BrScAs}, due to the
quasistructural nature \cite{20CsFaSa} of this molecular ion.

As to the utilization of quantum chemistry to solve the experimental puzzle,
through huge numerical efforts
%On the computational side, 
accurate rovibrational energy levels and eigenstates have been made available
for CH$_{5}^{+}$\  in recent years \cite{08WaCa,16WaCa,17FaQuCs}. 
These studies have revealed close-lying clusters in the rovibrational energy levels,
with fascinating symmetry characteristics. 
These features have defied explanation by conventional means, 
motivating the development of novel models for CH$_{5}^{+}$. 
The most important models put forward so far are as follows:
(a) particle-on-a-sphere (POS) \cite{84NaEzDeBe,86NaEzDeBe,88LeNaBeVi,88HuLeNaBe,05DeNe,08DeMcHuNe,14UhWaFo},
(b) five-dimensional (5D) rotor (superrotor) \cite{16ScJeSc,17ScJeSc,17ScJeSc2},
and (c) quantum graph \cite{18FaCs,19Rawlinso}.
So far, the quantum-graph model seems to have resulted in the most satisfactory explanation of the low-energy quantum dynamics of CH$_{5}^{+}$,
including both vibrations \cite{18FaCs} and rotations \cite{19Rawlinso}.

Quantum graphs have a long history in chemistry and physics,
dating back to Linus Pauling's description
of electrons in organic molecules in the 1930s \cite{36Pauling}. 
They have only recently been introduced to the study of nuclear dynamics,
where they have proved useful in high-resolution 
spectroscopy \cite{18FaCs,19Rawlinso} and also in explaining
$\alpha$-cluster dynamics in nuclear physics \cite{18Rawlinso,20Rawlinso}. 
Quantum graphs \cite{13BeKu} %model simply states that a quantum graph 
are metric graphs, that is each of their edges possesses a length.
In the context of rovibrational dynamics of molecules, each vertex of the graph represents a version \cite{06BuJe} of an equilibrium structure.
Depending on the nuclear permutation-inversion symmetry \cite{06BuJe} 
of the molecule of a given composition, even if the molecule has
a single minimum on a given potential energy surface it may possess a large
number of versions.
The vertices defined by the versions are connected by edges 
which represent collective internal
motions converting different versions into each other.
Once a quantum graph is set up, one constructs the 
one-dimensional (1D) Schr\"odinger equation for a particle confined 
to the graph and solves it to determine the energy levels and eigenstates\dag.
In this way, the complex multidimensional rovibrational quantum dynamics of a
polyatomic molecule is mapped onto the effective motion of a 1D particle
confined to a much simpler space.

%\begin{figure}[h]
%    \centering
%    \includegraphics[height=4cm]{gamma_120.pdf}
%    \caption{Illustration of the 3-regular quantum graph $\Gamma_{120}$.
    %Note that i
%    In this model of the quantum dynamics of CH$_{5}^{+}$\  there are
%    two distinct edge lengths, corresponding to internal rotations
%    of the H$_2$ unit by 60$^{\rm o}$ and the flip motion that 
%    exchanges a pair of protons between the H$_2$ and CH$_3^+$ units.}
%\label{gamma_120}
%\end{figure}

In the case of CH$_{5}^{+}$, the equilibrium structure, the only one found on its
ground electronic state, is composed of a H$_2$ unit
sitting on top of a CH$_3^+$ tripod, an arrangement with $C_s$ point-group symmetry. 
The five protons can be rearranged in 5!=120 ways,
generating 120 symmetry-equivalent versions.
These versions become the 120 vertices
of a quantum graph $\Gamma_{120}$ \cite{18FaCs}.%,
%illustrated in Fig. \ref{gamma_120}. 
There are two types of motion interconverting the 120 equivalent versions,
equivalent to scrambling the H atoms of CH$_{5}^{+}$:
the internal rotations of the H$_2$
unit by 60$^{\rm o}$ (both clockwise and counterclockwise),
and the flip motion that exchanges a pair of
protons between the H$_2$ and CH$_3^+$ units.
%(see Fig. \ref{c2v} for the local structure of $\Gamma_{120}$). 
The barriers to these motions on the potential energy
hypersurface of CH$_{5}^{+}$\ \cite{93ScKiSc} are known to be relatively low. It is plausible that the low-energy dynamics is dominated by motion along these particular paths, so that motions other than the internal rotation and flip motions can
be disregarded.
Thus, one can take these motions to correspond to the edges
of $\Gamma_{120}$. 
As one flip edge and two internal rotation edges are connected 
to each vertex of $\Gamma_{120}$, each vertex has a 
degree of three ($\Gamma_{120}$ is a 3-regular graph). %,
%see Fig. \ref{gamma_120}). 
Due to the nature of the underlying internal motions,
the 120 internal rotation and 60 flip edges are assigned effective lengths $L_\mathrm{rot}$ and $L_\mathrm{flip}$, respectively.

As shown before \cite{18FaCs,19Rawlinso},
the quantum graph $\Gamma_{120}$ reproduces the 
low-energy rovibrational energy levels of CH$_{5}^{+}$,
as well as of CD$_5^+$, remarkably well when 
optimized values of $L_\mathrm{flip}$ and $L_\mathrm{rot}$ are used \dag. %\cite{18FaCs,19Rawlinso}
%(see the ESI\dag ~for the numerical values of the edge length
%parameters and the vibrational energy levels of CH$_{5}^{+}$).
For instance, the $\Gamma_{120}$ model %vibrational energy levels 
perfectly reproduces the curious block structure 
(states occuring in groups of 15 and $30$, see Table~\ref{table:block}) 
of the vibrational eigenstates of CH$_{5}^{+}$, first noted in a variational study 
of Wang and Carrington \cite{16WaCa} and later confirmed in Ref. \citenum{17FaQuCs}.
%(see Table~\ref{table:block} for the symmetry characteristics of the first 60
%vibrational states of CH$_{5}^{+}$). 
% The  $\Gamma_{120}$ model captures the ordering 
% of these energy levels and, crucially, their symmetry characteristics. 
As seen in Table~\ref{table:block},
rovibrational eigenstates of CH$_{5}^{+}$\ are labelled by irreducible representations
(irreps) of the molecular symmetry (MS) group \cite{06BuJe}
$S_5^* = S_5 \times \{E,E^*\}$, 
generated by $S_5$ permutations of the five protons together with 
spatial inversion $E^*$ ($E$ denotes the identity operation).

\begin{table}[h]
\begin{center}
    \begin{tabular}{|c|c|}
 \hline
 Block 1 & Block 2 \\
\hline
$0-60$ cm$^{-1}$ (15,15) & $110-200$ cm$^{-1}$ (15,15)  \\ 
$A_1^+ \oplus G_1^+ \oplus H_1^+ \oplus H_2^+ \oplus$ &  $G_1^+ \oplus H_1^+ \oplus I^+ \oplus$   \\
$G_2^- \oplus H_2^- \oplus I^-$ & $A_2^- \oplus G_2^- \oplus H_1^- \oplus H_2^-$     \\
\hline
\end{tabular}
\end{center}

\small
\caption{\label{table:block}
The block structure characterizing the first 60 vibrational states of CH$_{5}^{+}$,
revealed in %taken from Wang and Carrington 
variational computations \cite{16WaCa,17FaQuCs}.
The numbers in parentheses give the total number of positive and negative parity states
within a block, reflecting the degeneracy of the states}

\end{table}

% \begin{figure}[b!]
% \noindent \begin{centering}
% \includegraphics[scale=0.2]{Figs/fig1.png}
% \par\end{centering}
% \caption{The block structure characterizing the  $J=0$ states of CH$_{5}^{+}$,
% revealed first in %taken from Wang and Carrington 
% the variational computations of Ref.~\cite{16WaCa}. \label{fig:carrington}}
% \end{figure}

Beyond the existence of blocks, in Table~\ref{table:block}
one can notice other clear symmetry relations for the first 60 quantum states.
%energy levels. 
%For example, a 
A comparison of the group-theoretic relation
\begin{equation}
\left(A_{1}^{+}\oplus G_{1}^{+}\oplus H_{1}^{+}\oplus H_{2}^{+}\right)\otimes A_{2}^{-}\simeq A_{2}^{-}\oplus G_{2}^{-}\oplus H_{2}^{-}\oplus H_{1}^{-}
\label{eq1}
\end{equation}
with the data in Table~\ref{table:block} suggests a  direct correspondence 
between the 15 positive-parity states in Block $1$ 
[appearing on the left-hand side (LHS) of Eq.~(\ref{eq1})] and the 15 negative-parity states in Block $2$ [right-hand side (RHS) of Eq.~(\ref{eq1})].
Likewise, 
\begin{equation}
\left(G_{2}^{-}\oplus H_{2}^{-}\oplus I^{-}\right)\otimes A_{2}^{-} \simeq  G_{1}^{+}\oplus H_{1}^{+}\oplus I^{+},
\end{equation}
suggesting a link between the 15 negative-parity states in Block $1$ and
the positive-parity states in Block $2$. 
These remarkable symmetry relations have been lacking any simple explanation, 
even in terms of the $\Gamma_{120}$ model.
As this %One purpose of this 
paper proves, introduction of the simplest quantum-graph model, 
$\Gamma_{60}$, of the quantum dynamics of CH$_{5}^{+}$,
derived from $\Gamma_{120}$, %which 
is sufficient to explain the curious energy-level and symmetry structure 
of the lowest vibrational states of CH$_{5}^{+}$, and, as a bonus feature, it allows the
analytic determination of the quantum states of the model problem.

%\section*{The 60-vertex model}
% \subsection{Setting up the model}}
%Our starting point is the quantum graph $\Gamma_{120}$ that 
%was successfully used by us to describe the low-energy rovibrational 
%quantum dynamics of CH$_{5}^{+}$\  \cite{18FaCs,19Rawlinso}.
%The core idea of the quantum-graph treatment of CH$_{5}^{+}$\ is that 
%the 120 symmetry-equivalent versions of CH$_{5}^{+}$\ are assigned to the 
%120 vertices of $\Gamma_{120}$.
%We assume that only the internal rotation and flip motions of CH$_{5}^{+}$\ are
%of relevance %in our case 
%and the vertices of $\Gamma_{120}$ are connected by 
%internal rotation or flip edges if the corresponding versions %can be
%interconverted by an internal rotation or a flip motion, respectively.

Let us start our journey toward the simplest model
with the quantum graph $\Gamma_{120}$. 
We recall two important characteristics of our original study \cite{18FaCs}.
First, we neglect the potential energy
along the edges of $\Gamma_{120}$, since the barriers 
to the internal rotation and flip motions are small 
(about $30$ cm$^{-1}$ and $300$ cm$^{-1}$, respectively \cite{93ScKiSc}). 
Second, we fix the effective edge lengths $L_\mathrm{flip}$ and $L_\mathrm{rot}$.
In Ref. \citenum{18FaCs} this was done by an optimization procedure to give the best fit to either 7D or 12D reference data. 
In both cases the optimized $L_\mathrm{flip}$ was much smaller than the optimized $L_\mathrm{rot}$, with the ratio  %$L_\mathrm{flip}/L_\mathrm{rot} = \approx 1/60$
$L_\mathrm{flip}/L_\mathrm{rot} = 1.0 / 61.2$
in the 7D case.

%\red{Hint: the figures with the CH$_{5}^{+}$\ structures and graphs should be embedded here.}

\begin{figure}[h]
    \centering
\includegraphics[height=6cm]{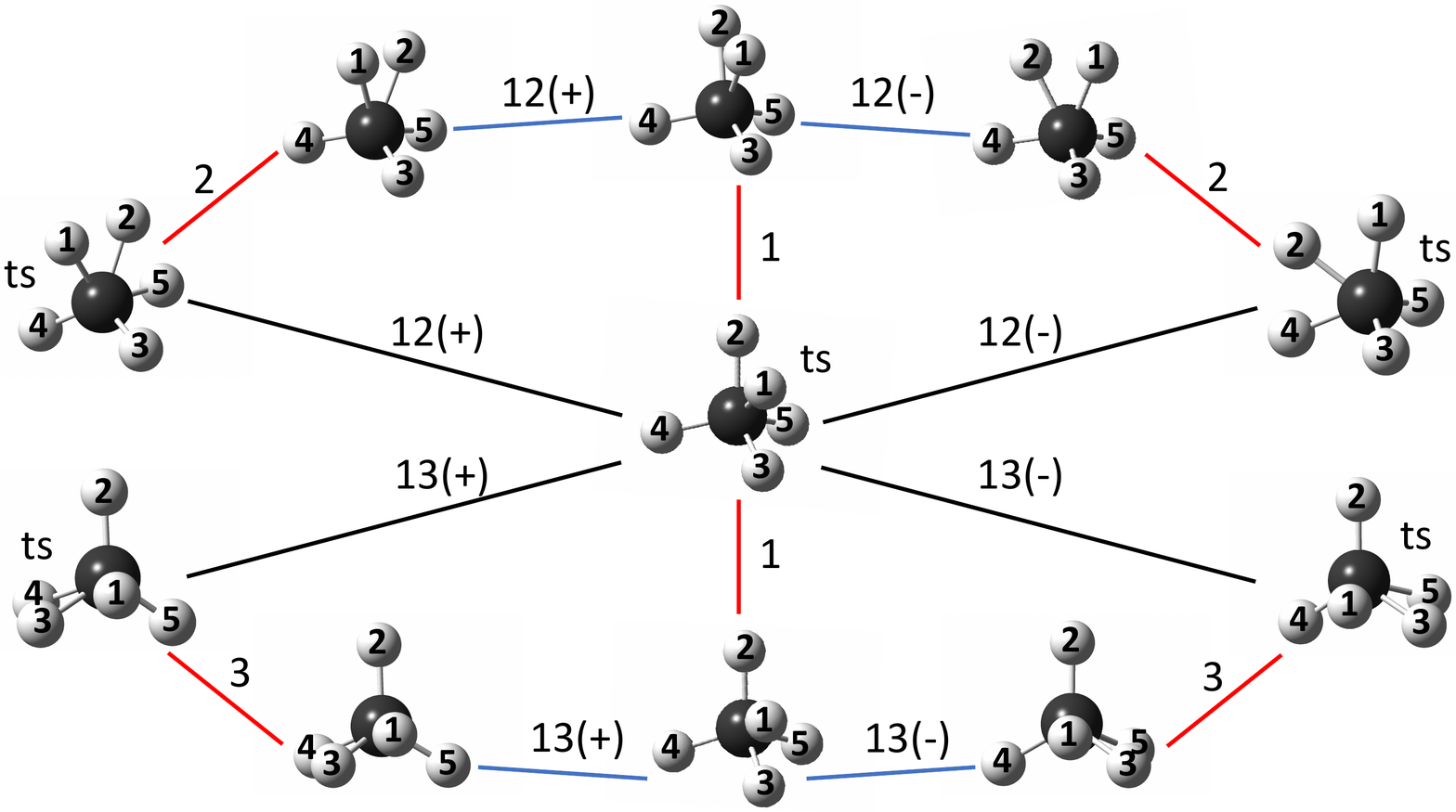}
\caption{Local structure of the quantum graphs $\Gamma_{120}$
(blue and red edges) and $\Gamma_{60}$ (black edges). 
The red edges correspond to the flip motion and the labels indicate which proton is exchanged from a H$_2$ unit to a CH$_3^+$ unit. 
The blue edges correspond to an internal rotation and the labels indicate the H$_2$ unit which rotates relative to the CH$_3^+$ unit in a clockwise ($+$) or anticlockwise ($-$) fashion. The midpoint of each red flip edge is a $C_{2v}$-symmetric transition state (ts). In going from $\Gamma_{120}$ to $\Gamma_{60}$, the red edges shrink so that we are left with just the transition states connected by black edges.}
\label{c2v}
\end{figure}

\begin{figure}[h]
    \centering
    \includegraphics[height=6cm]{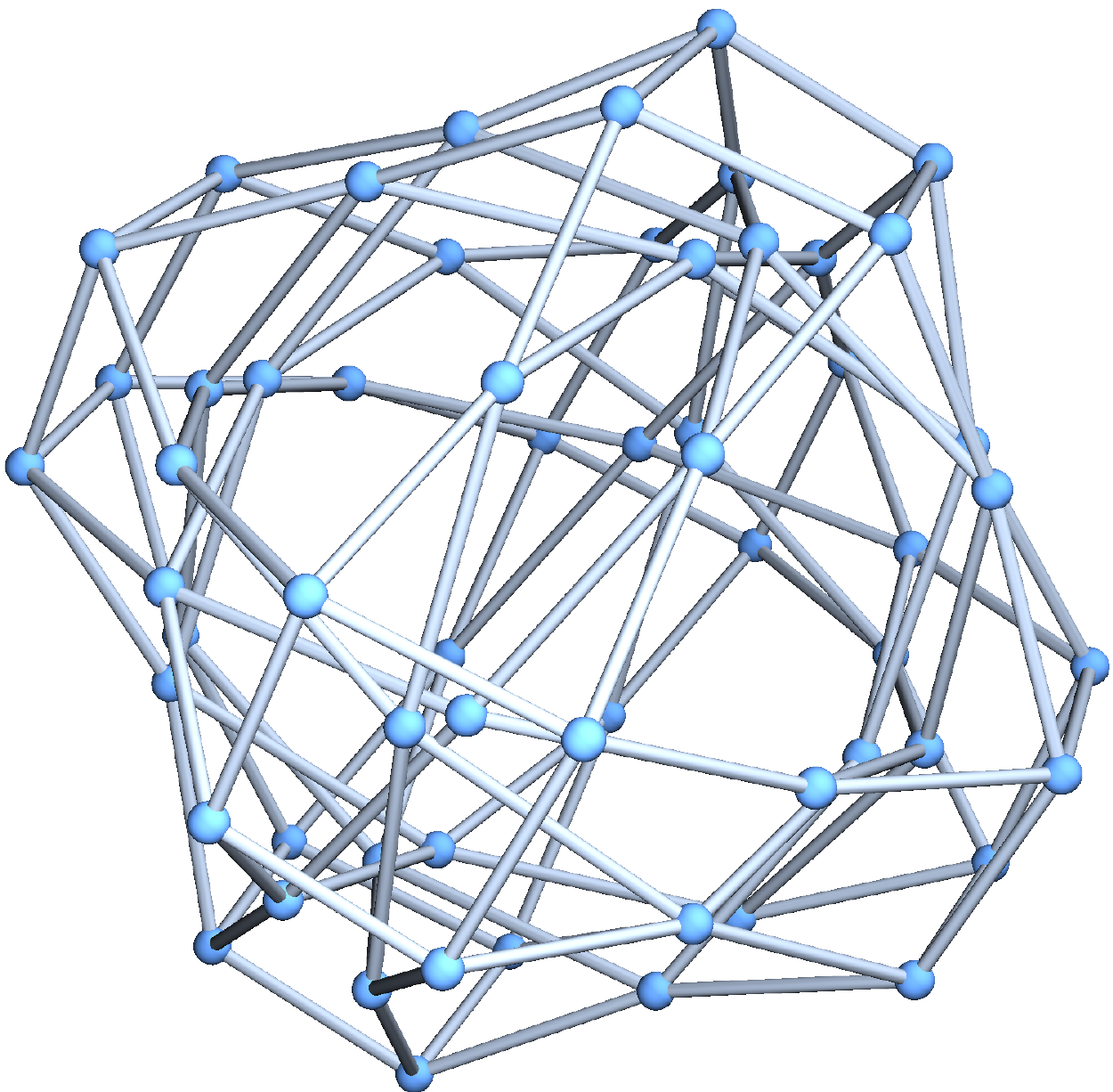}
    \caption{Illustration of the 4-regular quantum graph $\Gamma_{60}$.
    In this model of the quantum dynamics of CH$_{5}^{+}$\  there is a single
    edge length, connecting versions of $C_{2v}$-symmetric transition states,
    midpoints of the flip edge of $\Gamma_{120}$.}
\label{gamma_60}
\end{figure}

Our new model is based on the following idea: the ratio $L_\mathrm{flip}/L_\mathrm{rot}$ is so small that it is tempting to imagine shrinking the flip edges to zero length, identifying the two vertices at the endpoints of each flip edge to give a single vertex. Setting $L_\mathrm{flip}=0$ has a negligible effect on the accuracy of the fit, at least at low energies. At the same time, this approximation gives a huge simplification: the number of vertices is halved and we get a new quantum graph, $\Gamma_{60}$, with only the internal rotation edges remaining. 
It is reasonable to identify each new vertex with the midpoint of the (now contracted) flip edge, which is a $C_{2v}$-symmetric transition state, as illustrated in Fig. \ref{c2v}.
$\Gamma_{60}$ represents $60$ symmetry-equivalent versions
of this configuration. %arrangement,
We propose that the most important characteristics of the
low-energy quantum states of CH$_{5}^{+}$\  can be understood
in terms of a 1D, potential-free motion between these versions
corresponding to the vertices of the quantum graph $\Gamma_{60}$. 
Note that each vertex is connected to precisely four other vertices,
as shown also in Fig. \ref{c2v},
giving rise to the 4-regular (quartic) quantum graph $\Gamma_{60}$, illustrated in Fig. \ref{gamma_60}. 

%One way of thinking of
%$\Gamma_{60}$ %this graph
%is that it is the graph obtained
%if one starts with $\Gamma_{120}$ and %shrinks the edge length 
%corresponding to flip motions to zero length. 
%Indeed, this model was inspired by %that work as in
%the fact that in Ref.~\cite{18FaCs}
%their modelling they actually take 
%the flip-motion edge length was set to
%be much smaller than the internal-rotation edge %length.
There is an alternative way of rationalizing the above contraction procedure.
%is as follows:
At the energies we are interested in, one can show that the $\Gamma_{120}$ wave functions for the energy eigenstates are approximately constant
along the flip edges. 
In this limit, the boundary
conditions of $\Gamma_{120}$ become equivalent
to those of $\Gamma_{60}$\dag.
Either way,  $\Gamma_{60}$ only retains edges 
corresponding to the internal rotation. 
Our simplified model therefore has the feature of explaining the 
low-energy dynamics solely in terms of the internal rotation motion 
without the flip motion, with the constant wave function argument 
allowing for backstage full exchange of the protons. 
This model is thus set up in clear violation of the claim of the authors 
of Ref. \citenum{13WiIvMa}, namely that 
``the combination of the two [internal motions]
enables large-amplitude motion and thus 
''full scrambling`` ... whereas partial scrambling leads to
the well-known small-amplitude motion only''.

%The quantum graph $\Gamma_{60}$ 
%generated by these motions 
%has $60$ vertices and $120$ edges. 
%Each symmetry-equivalent vertex $v_{i}\in V$ corresponds
%to a version of the $C_{2v}$-symmetric arrangement and each of the
%symmetry-equivalent edges joins two such arrangements. 
%The action
%of the full molecular symmetry (MS) group $\mathbb{Z}_{2} \otimes S_{5}$,
%generated by inversion and by permutations of the %five identical protons,
%acts on %the graph 
%$\Gamma_{60}$ by graph automorphisms. \\

%\subsection{Finding energy levels for $\Gamma_{60}$}
We now seek the quantum states corresponding to motion on 
the $\Gamma_{60}$ graph. 
The eigenenergies are found by solving the time-independent 
Schr\"odinger equation for a free particle moving along the edges, 
with the so-called Neumann boundary conditions \cite{13BeKu} 
imposed on the eigenstates. 
These conditions are that the wave function should be continuous everywhere,
with zero total momentum flux out of each vertex. 
As we have already pointed out, $\Gamma_{60}$ is a 4-regular graph 
with all edges having a common length $l=L_\mathrm{rot}$. 
Perhaps surprisingly, these properties imply that the structure 
of the quantum energy levels can be determined entirely from \emph{combinatorial properties} of the graph.

More precisely, given a wave function $\psi$ defined on the graph $\Gamma_{60}$ 
and obeying the time-independent Schr\"odinger equation along each edge,
\begin{equation}
-\frac{1}{2}\frac{\textrm{d}^{2} \psi}{\textrm{d}x^{2}}=E\psi,
\end{equation}
where $x$ is a mass-scaled coordinate, consider the vector of its values at each vertex 
$\mathbf{v}=\left(\psi\left(v_{1}\right),\psi\left(v_{2}\right),\ldots\right)$.
It is straightforward to prove\dag %, see the ESI\dag~,
 that $\psi$ is an %energy 
eigenfunction with energy $E$ satisfying the Neumann boundary conditions 
if and only if
\begin{equation}
\bm{A}\mathbf{v}=4\cos\left(\sqrt{2E}l\right)\mathbf{v},
\label{sp}
\end{equation}
\emph{i.e.}, if and only if $\lambda=4\cos\left(\sqrt{2E}l\right)$ 
is an eigenvalue of the \emph{adjacency matrix} $\bm{A}$ %(defined below)
for the graph $\Gamma_{60}$,
with $\mathbf{v}$ in the corresponding eigenspace. 
% The matrix $\bm{A}$ on the LHS is the \emph{adjacency matrix}
% for the graph $\Gamma_{60}$. 
$\bm{A}$ is simply a matrix whose elements indicate
whether given pairs of vertices are connected by an edge or not:
\begin{equation}
\left(A\right)_{ij}=\begin{cases}
1 & \mathrm{if~vertices~}\text{}v_{i},v_{j}\text{}\mathrm{~connected}\\
0 & \mathrm{otherwise}
\end{cases}
\end{equation}
and is a familiar concept in elementary graph theory \cite{93Bi}. 
%$\Gamma_{60}$ is associated with a $60\times60$ adjacency matrix 
%$\bm{A}\in\mathrm{Mat}_{60\times60}$
%Its entries are given by 
%The factor of $4$ on the RHS of Eq.~(\ref{sp}) corresponds to
%the fact that each vertex connects to precisely $4$ other vertices. 

Equation (\ref{sp}) therefore relates the  \emph{quantum spectrum}
(the eigenvalues of the Hamiltonian) to the so-called 
\emph{combinatorial spectrum} (the eigenvalues of the adjacency matrix). 
The combinatorial spectrum is
a concept already utilized in molecular spectroscopy \cite{16ArFuCs}, 
and only depends on the connectivity of the graph as encoded in $\bm{A}$.

%\section*{Results and discussion}

To find the combinatorial spectrum of $\Gamma_{60}$,
we look for roots of the characteristic polynomial 
$\chi_{\bm{A}}\left(\lambda\right)=\mathrm{det}\left(\lambda \bm{I}-\bm{A}\right)$
associated with the adjacency matrix $\bm{A}$. 
An explicit expression for $\bm{A}$ is easily derived by considering
paths of the form illustrated in Fig.~\ref{c2v}. 
In the end, we obtain
\begin{equation}
\label{eq:lambda}
\chi_{\bm{A}}\left(\lambda\right)=\left(\lambda^{4}-9\lambda^{2}+16\right)^{5}\left(\lambda^{4}-12\lambda^{2}+16\right)^{4} \left(\lambda^{2}-1\right)^{11}\left(\lambda^{2}-16\right),
\end{equation}
and the full combinatorial spectrum is given in Table~\ref{spec}.
Table~\ref{spec} also shows the dimensions of the corresponding eigenspaces 
and the irreps of the MS group $S_5^*$.

% original version of the table
%\begin{table}
%\begin{centering}
%%\begin{tabular}{|c|c|c|}
%\begin{tabular}{ccc}
%\hline \hline
%$\lambda$ & $\mathrm{dim}\left(E_{\lambda}\right)$ & $\mathbb{Z}_{2}\otimes S_{5}$ irrep\tabularnewline
%\hline 
%$4$ & $1$ & $A_{1}^{+}$\tabularnewline
%%\hline 
%$1+\sqrt{5}$ & $4$ & $G_{2}^{-}$\tabularnewline
%%\hline 
%$\frac{1}{2}\left(1+\sqrt{17}\right)$ & $5$ & $H_{1}^{+}$\tabularnewline
%%\hline 
%$\frac{1}{2}\left(-1+\sqrt{17}\right)$ & $5$ & $H_{2}^{-}$\tabularnewline
%%\hline 
%$-1+\sqrt{5}$ & $4$ & $G_{1}^{+}$\tabularnewline
%%\hline 
%$1$ & $11$ & $H_{2}^{+}\oplus I^{-}$\tabularnewline
%%\hline 
%$-1$ & $11$ & $H_{1}^{-}\oplus I^{+}$\tabularnewline
%%\hline 
%$1-\sqrt{5}$ & $4$ & $G_{2}^{-}$\tabularnewline
%%\hline 
%$\frac{1}{2}\left(1-\sqrt{17}\right)$ & $5$ & $H_{1}^{+}$\tabularnewline
%%\hline 
%$\frac{1}{2}\left(-1-\sqrt{17}\right)$ & $5$ & $H_{2}^{-}$\tabularnewline
%%\hline 
%$-1-\sqrt{5}$ & $4$ & $G_{1}^{+}$\tabularnewline
%%\hline 
%$-4$ & $1$ & $A_{2}^{-}$\tabularnewline
%\hline \hline
%\end{tabular}
%\par\end{centering}
%\caption{Spectrum of the quantum graph $\Gamma_{60}$. \label{spec}}
%\end{table}

\begin{table}[h]
\begin{center}
\begin{tabular}{|c|c|c|}
\hline 
$\lambda$ & $\mathrm{dim}\left(\lambda\right)$ & $S_5^*$ irrep \\
\hline
$4$ & $1$ & $A_{1}^{+}$ \\ 
$1+\sqrt{5}$ & $4$ & $G_{2}^{-}$ \\ 
$\frac{1}{2}\left(1+\sqrt{17}\right)$ & $5$ & $H_{1}^{+}$ \\
$\frac{1}{2}\left(-1+\sqrt{17}\right)$ & $5$ & $H_{2}^{-}$ \\
$-1+\sqrt{5}$ & $4$ & $G_{1}^{+}$ \\
$1$ & $11$ & $H_{2}^{+}\oplus I^{-}$ \\
$-1$ & $11$ & $H_{1}^{-}\oplus I^{+}$ \\
$1-\sqrt{5}$ & $4$ & $G_{2}^{-}$ \\
$\frac{1}{2}\left(1-\sqrt{17}\right)$ & $5$ & $H_{1}^{+}$ \\
$\frac{1}{2}\left(-1-\sqrt{17}\right)$ & $5$ & $H_{2}^{-}$ \\
$-1-\sqrt{5}$ & $4$ & $G_{1}^{+}$ \\
$-4$ & $1$ & $A_{2}^{-}$ \\
\hline
\end{tabular}
\end{center}
\small
\caption{\label{spec}The combinatorial spectrum of the quantum graph $\Gamma_{60}$, where $\mathrm{dim}\left(\lambda\right)$ gives the degeneracy of a given eigenvector
corresponding to the eigenvalue $\lambda$ [see Eq.~(\ref{eq:lambda})]}

\label{tablecomb}
\end{table}

%\subsection{Consequences}
We pause here to note the striking similarity between Tables~\ref{table:block} and \ref{spec}. 
%s of Wang and Carrington \cite{16WaCa} (as reproduced in Fig. \ref{fig:carrington}). 
First, note that the combinatorial 
spectrum splits into positive $\lambda$ and negative $\lambda$,
with each corresponding to a total eigenspace dimension of $30$.
Moreover, the eigenspaces associated with positive $\lambda$ transform
in precisely the same irreps as %Wang and Carrington's 
Block $1$ of Table~\ref{table:block},
while those associated with negative $\lambda$ transform precisely like 
%Wang and Carrington's 
Block $2$. 
Thus, purely combinatorial properties of the quantum graph $\Gamma_{60}$
have captured the block structure of the lowest vibrational states of CH$_{5}^{+}$.
Even more interestingly, we have an explanation for the curious
relationship between Block $1$ states and Block $2$ states:
this corresponds to a  $\lambda\rightarrow-\lambda$ symmetry of the combinatorial spectrum (see Table~\ref{tablecomb}), under which the $S_5^*$ irreps are related by multiplication with $A_2^-$. The symmetry of the combinatorial spectrum under $\lambda\rightarrow-\lambda$ is a simple consequence \cite{93Bi}
of the fact that the quantum graph $\Gamma_{60}$
is \emph{bipartite}:
the set of vertices $V$ can be divided into two disjoint and independent
sets $A$ and $B$ such that every edge connects a vertex in $A$ to one in $B$. 
The sets $A$ and $B$ are related by odd permutations of the protons\dag. %A detailed explanation for the relationship between $S_5^*$ irreps under this symmetry is given in section IV of the ESI\dag.

\begin{figure}[h]
    \centering
%\noindent \begin{centering}
%\includegraphics[scale=0.36]{Figs/fig3}
%\par\end{centering}
\includegraphics[height=6cm]{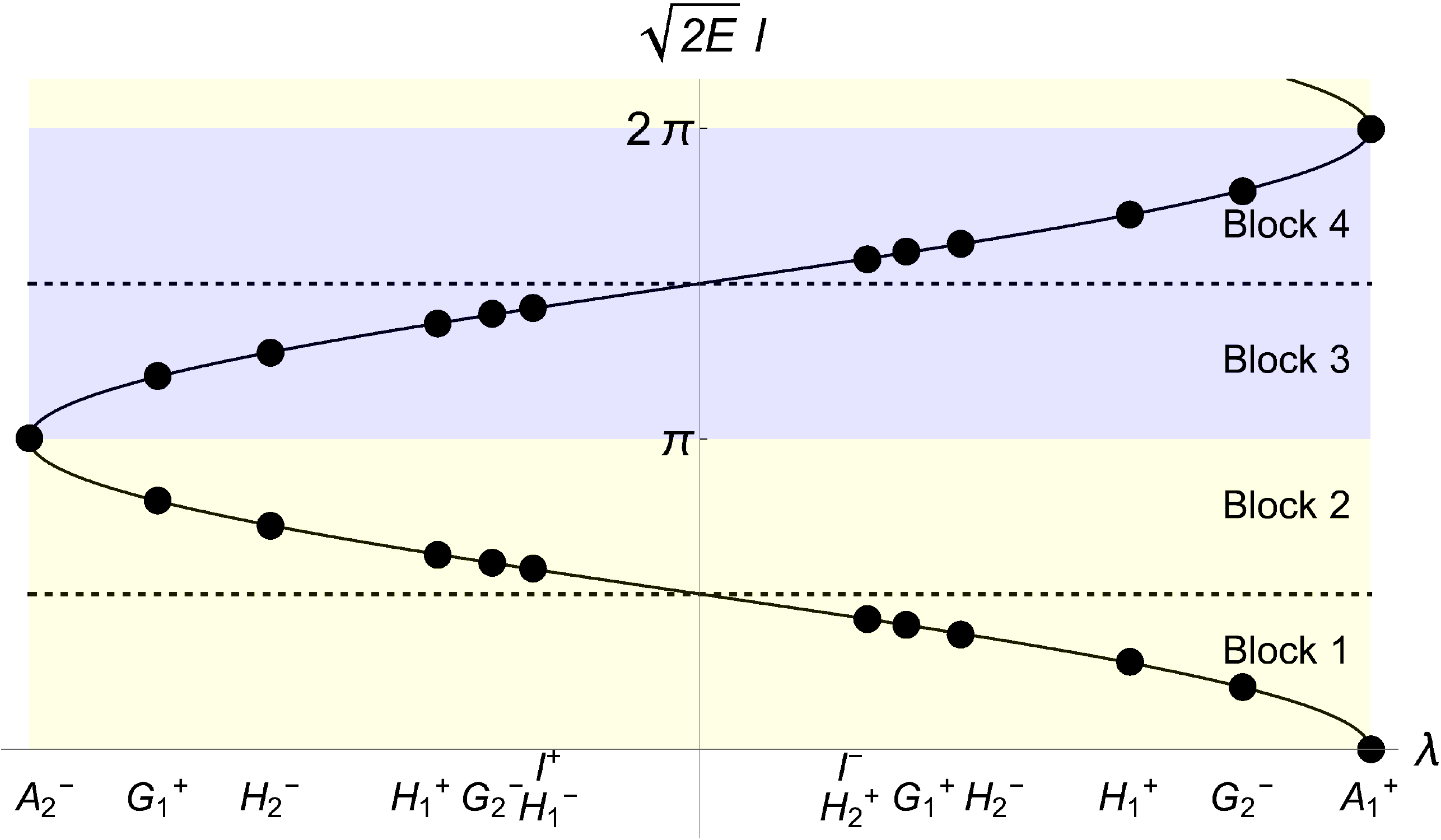}
\caption{Illustration of the block structure and the symmetry properties
of the spectrum of the quantum graph $\Gamma_{60}$.
Black dots indicate energies of the quantum states.}
\label{quantumspec}
\end{figure}

Equation (\ref{sp}) relates the combinatorial spectrum to the quantum spectrum,
as illustrated in Fig. \ref{quantumspec}. 
We can see the consequences of the $\lambda\rightarrow-\lambda$ symmetry for the quantum energy levels: each state in Block $1$ comes 
with a partner in Block $2$, with their corresponding
values of $\sqrt{2E}l$ being related by reflection in the line $\sqrt{2E}l=\pi/2$.
In particular, the dimensionless ratios
\begin{equation}
\frac{\sqrt{E_{1}\left(I^{-}\right)}+\sqrt{E_{2}\left(I^{+}\right)}}{\sqrt{E_{1}\left(H_{1}^{+}\right)}+\sqrt{E_{2}\left(H_{2}^{-}\right)}},\frac{\sqrt{E_{1}\left(H_{2}^{+}\right)}+\sqrt{E_{2}\left(H_{1}^{-}\right)}}{\sqrt{E_{1}\left(H_{1}^{+}\right)}+\sqrt{E_{2}\left(H_{2}^{-}\right)}},\ldots
\end{equation}
%\begin{multline}
% \sqrt{E_{\mathrm{Block}1}\left(I^{-}\right)}+\sqrt{E_{\mathrm{Block}2}\left(I^{+}\right)}=\sqrt{E_{\mathrm{Block}1}\left(H_{2}^{+}\right)}+\sqrt{E_{\mathrm{Block}2}\left(H_{1}^{-}\right)}=\ldots=\frac{\pi}{\sqrt{2}l}.
%\sqrt{E_{1}\left(I^{-}\right)}+\sqrt{E_{2}\left(I^{+}\right)}=\sqrt{E_{1}\left(H_{2}^{+}\right)}+\sqrt{E_{2}\left(H_{1}^{-}\right)} \\ =\ldots=\frac{\pi}{\sqrt{2}l},
%\end{multline}
%where the subscripts 1 and 2 stand for Block 1 and Block %2, respectively.
%\begin{equation}
%\frac{\sqrt{E_{\mathrm{Block}1}\left(I^{-}\right)}+\sqrt{E_{\mathrm{Block}2}\left(I^{+}\right)}}{\sqrt{E_{\mathrm{Block}1}\left(H_{1}^{+}\right)}+\sqrt{E_{\mathrm{Block}2}\left(H_{2}^{-}\right)}},\frac{\sqrt{E_{\mathrm{Block}1}\left(H_{2}^{+}\right)}+\sqrt{E_{\mathrm{Block}2}\left(H_{1}^{-}\right)}}{\sqrt{E_{\mathrm{Block}1}\left(H_{1}^{+}\right)}+\sqrt{E_{\mathrm{Block}2}\left(H_{2}^{-}\right)}},\ldots
%\end{equation}
are all equal to $1$ in the $\Gamma_{60}$ model.
This compares very favourably with the variational seven-dimensional
model \cite{08WaCa,16WaCa,17FaQuCs} results
%Wang and Carrington \cite{16WaCa} values of
\begin{equation}
\frac{\sqrt{E_{1}\left(A_{1}^{+}\right)}+\sqrt{E_{2}\left(A_{2}^{-}\right)}}{\sqrt{E_{1}\left(H_{1}^{+}\right)}+\sqrt{E_{2}\left(H_{2}^{-}\right)}}\approx\frac{\sqrt{0}+\sqrt{198}}{\sqrt{20}+\sqrt{139}}\approx0.86,
% \frac{\sqrt{E_{\mathrm{Block\,}1}\left(A_{1}^{+}\right)}+\sqrt{E_{\mathrm{Block\,}2}\left(A_{2}^{-}\right)}}{\sqrt{E_{\mathrm{Block\,}1}\left(H_{1}^{+}\right)}+\sqrt{E_{\mathrm{Block\,}2}\left(H_{2}^{-}\right)}}\approx\frac{\sqrt{0.0}+\sqrt{198.1}}{\sqrt{20.4}+\sqrt{139.4}}\approx0.86
\end{equation}
\begin{equation}
\frac{\sqrt{E_{1}\left(G_{2}^{-}\right)}+\sqrt{E_{2}\left(G_{1}^{+}\right)}}{\sqrt{E_{1}\left(H_{1}^{+}\right)}+\sqrt{E_{2}\left(H_{2}^{-}\right)}}\approx\frac{\sqrt{10}+\sqrt{154}}{\sqrt{20}+\sqrt{139}}\approx0.95,
\end{equation}
%\begin{equation}
%\frac{\sqrt{E_{\mathrm{Block\,}1}\left(G_{2}^{-}\right)}+\sqrt{E_{\mathrm{Block\,}2}\left(G_{1}^{+}\right)}}{\sqrt{E_{\mathrm{Block\,}1}\left(H_{1}^{+}\right)}+\sqrt{E_{\mathrm{Block\,}2}\left(H_{2}^{-}\right)}}\approx\frac{\sqrt{9.9}+\sqrt{154.5}}{\sqrt{20.4}+\sqrt{139.4}}\approx0.95
%\end{equation}
\begin{equation}
\frac{\sqrt{E_{1}\left(H_{2}^{-}\right)}+\sqrt{E_{2}\left(H_{1}^{+}\right)}}{\sqrt{E_{1}\left(H_{1}^{+}\right)}+\sqrt{E_{2}\left(H_{2}^{-}\right)}}\approx\frac{\sqrt{41}+\sqrt{122}}{\sqrt{20}+\sqrt{139}}\approx1.07,
\end{equation}
%\begin{equation}
%\frac{\sqrt{E_{\mathrm{Block}1}\left(H_{2}^{-}\right)}+\sqrt{E_{\mathrm{Block}2}\left(H_{1}^{+}\right)}}{\sqrt{E_{\mathrm{Block}1}\left(H_{1}^{+}\right)}+\sqrt{E_{\mathrm{Block}2}\left(H_{2}^{-}\right)}}\approx\frac{\sqrt{41.1}+\sqrt{122.0}}{\sqrt{20.4}+\sqrt{139.4}}\approx1.07
%\end{equation}
\begin{equation}
\frac{\sqrt{E_{1}\left(G_{1}^{+}\right)}+\sqrt{E_{2}\left(G_{2}^{-}\right)}}{\sqrt{E_{1}\left(H_{1}^{+}\right)}+\sqrt{E_{2}\left(H_{2}^{-}\right)}}\approx\frac{\sqrt{49}+\sqrt{113}}{\sqrt{20}+\sqrt{139}}\approx1.08,
\end{equation}
%\begin{equation}
%\frac{\sqrt{E_{\mathrm{Block}1}\left(G_{1}^{+}\right)}+\sqrt{E_{\mathrm{Block}2}\left(G_{2}^{-}\right)}}{\sqrt{E_{\mathrm{Block}1}\left(H_{1}^{+}\right)}+\sqrt{E_{\mathrm{Block}2}\left(H_{2}^{-}\right)}}\approx\frac{\sqrt{49.4}+\sqrt{112.7}}{\sqrt{20.4}+\sqrt{139.4}}\approx1.08
%\end{equation}
\begin{equation}
\frac{\sqrt{E_{1}\left(I^{-}\right)}+\sqrt{E_{2}\left(I^{+}\right)}}{\sqrt{E_{1}\left(H_{1}^{+}\right)}+\sqrt{E_{2}\left(H_{2}^{-}\right)}}\approx\frac{\sqrt{58}+\sqrt{112}}{\sqrt{20}+\sqrt{139}}\approx1.12,
\end{equation}
%\begin{equation}
%\frac{\sqrt{E_{\mathrm{Block}1}\left(I^{-}\right)}+\sqrt{E_{\mathrm{Block}2}\left(I^{+}\right)}}{\sqrt{E_{\mathrm{Block}1}\left(H_{1}^{+}\right)}+\sqrt{E_{\mathrm{Block}2}\left(H_{2}^{-}\right)}}\approx\frac{\sqrt{58.4}+\sqrt{112.0}}{\sqrt{20.4}+\sqrt{139.4}}\approx1.12
%\end{equation}
and
\begin{equation}
\frac{\sqrt{E_{1}\left(H_{2}^{+}\right)}+\sqrt{E_{2}\left(H_{1}^{-}\right)}}{\sqrt{E_{1}\left(H_{1}^{+}\right)}+\sqrt{E_{2}\left(H_{2}^{-}\right)}}\approx\frac{\sqrt{59}+\sqrt{114}}{\sqrt{20}+\sqrt{139}}\approx1.12.
\end{equation}
%\begin{equation}
%\frac{\sqrt{E_{\mathrm{Block}1}\left(H_{2}^{+}\right)}+\sqrt{E_{\mathrm{Block}2}\left(H_{1}^{-}\right)}}{\sqrt{E_{\mathrm{Block}1}\left(H_{1}^{+}\right)}+\sqrt{E_{\mathrm{Block}2}\left(H_{2}^{-}\right)}}\approx\frac{\sqrt{59.3}+\sqrt{113.7}}{\sqrt{20.4}+\sqrt{139.4}}\approx1.12.
%\end{equation}

%\section*{Conclusions}

In this paper we have drastically simplified the quantum graph model of the
low-energy rovibrational quantum dynamics of CH$_{5}^{+}$\ by reducing the original 
120-vertex quantum graph to a 60-vertex graph, $\Gamma_{60}$.
$\Gamma_{60}$ was constructed by shrinking the edges 
corresponding to the flip internal motion that exchanges a pair of
protons between the H$_2$ and CH$_3^+$ units of the equilibrium structure of CH$_{5}^{+}$.
Thus, at first sight we neglect one of the two important 
large-amplitude internal motions
characterizing the exchange dynamics (scrambling) of the H atoms of CH$_{5}^{+}$.
%It is even more surprising that we keep the internal rotation of the H$_2$
%molecule of the equilibrium structure, sitting on top of the CH$_3^+$ tripod.
This allows us to obtain the quantum states of $\Gamma_{60}$
%are obtained in 
in analytic form, with the structure of the energy levels depending only 
on combinatorial properties.
The eigenvalues of this simple 1D, potential-free model are
in excellent agreement with the first 60 vibrational states
determined by sophisticated variational
nuclear-motion computations utilizing a potential energy hypersurface.
Furthermore, the bipartite structure of $\Gamma_{60}$ 
gives a natural explanation for symmetries
in the vibrational energy level structure of CH$_{5}^{+}$, 
again in perfect agreement with the results of 
variational nuclear dynamics computations.
Note that neither the variational %quantum-chemical 
computations \cite{08WaCa,16WaCa,17FaQuCs} nor the quantum-graph models \cite{18FaCs,19Rawlinso}
yield only the Pauli-allowed states of CH$_{5}^{+}$\ (states with $A_2^{\pm}$, 
$G_2^{\pm}$, and $H_2^{\pm}$ symmetry have non-zero spin-statistical weights),
so our discussion focused on \emph{all}
possible states; the non-existing states can be filtered out \emph{a posteriori}.

\section*{Acknowledgements}
The work of JIR was supported by the EPSRC grant CHAMPS EP/P021123/1.
The work performed in Budapest received support from NKFIH 
(grant no. K119658) and
from the  ELTE Institutional Excellence Program (TKP2020-IKA-05) 
financed by the Hungarian Ministry of Human Capacities.
%%%END OF MAIN TEXT%%%

%  For footnotes in the main text of the article please number the footnotes to avoid duplicate symbols. e.g.  \footnote[num]{your text} the corresponding author \ast counts as footnote 1, ESI as footnote 2, e.g. if there is no ESI, please start at [num]=[2], if ESI is cited in the title please start at [num]=[3] etc. Please also cite the ESI within the main body of the text using \dag.

% The \balance command can be used to balance the columns on the final page if desired. It should be placed anywhere within the first column of the last page.

% \balance

% If notes are included in your references you can change the title from 'References' to 'Notes and references' using the following command:
% \renewcommand\refname{Notes and references}

%%%REFERENCES%%%
\bibliographystyle{unsrt}
\bibliography{rsc}

\begin{thebibliography}{10}

\bibitem{68OlSc}
George~A. Olah and Richard~H. Schlosberg.
\newblock Chemistry in super acids. i. hydrogen exchange and polycondensation
  of methane and alkanes in fso$_3$h-sbf$_5$ ("magic acid") solution.
  protonation of alkanes and the intermediacy of ch$_5^+$ and related
  hydrocarbon ions. the high chemical reactivity of "paraffins" in ionic
  solution reactions.
\newblock {\em J. Am. Chem. Soc.}, 90(10):2726--2727, 1968.

\bibitem{69OlKlSc}
George~A. Olah, Gilles Klopman, and Richard~H. Schlosberg.
\newblock Super acids. iii. protonation of alkanes and intermediacy of
  alkanonium ions, pentacoordinated carbon cations of ch$_5^+$ type. hydrogen
  exchange, protolytic cleavage, hydrogen abstraction; polycondensation of
  methane, ethane, 2,2-dimethylpropane and 2,2,3,3-tetramethylbutane in
  fso$_3$h-sbf$_5$.
\newblock {\em J. Am. Chem. Soc.}, 91(12):3261--3268, 1969.

\bibitem{74Olah}
G.~A. Olah.
\newblock {\em Carbocations and Electrophilic Reactions}.
\newblock VCH-Wiley Publishers, Weinheim, 1974.

\bibitem{94Olah}
G.~A. Olah.
\newblock {\em My Search for Carbocations and Their Role in Chemistry}, pages
  149--176.
\newblock December 1994.

\bibitem{99WhTaOk}
Edmund~T. White, Jiang Tang, and Takeshi Oka.
\newblock {CH$_5^+$: the infrared spectrum observed}.
\newblock {\em Science}, 284:135--137, 1999.

\bibitem{05AsKuRe}
Oskar Asvany, Padmanabhan Padma~Kumar, Britta Redlich, Ilka Hegemann, Stephan
  Schlemmer, and Dominik Marx.
\newblock Understanding the infrared spectrum of bare ch$_5^+$.
\newblock {\em Science}, 309(5738):1219--1222, 2005.

\bibitem{08WaCa}
X.-G. Wang and T.~{Carrington Jr.}
\newblock {Vibrational energy levels of CH$_5^+$}.
\newblock {\em J. Chem. Phys.}, 129:234102, 2008.

\bibitem{10IvAsWiHu}
Sergei~D. Ivanov, Oskar Asvany, Alexander Witt, Edouard Hugo, Gerald Mathias,
  Britta Redlich, Dominik Marx, and Stephan Schlemmer.
\newblock {Quantum-induced symmetry breaking explains infrared spectra of
  CH$_5^+$ isotopologues}.
\newblock {\em Nat. Chem.}, 2:298--302, 2010.

\bibitem{15AsYaBrPo}
Oskar Asvany, Koichi M.~T. Yamada, Sandra Br\"unken, Alexey Potapov, and
  Stephan Schlemmer.
\newblock {Experimental ground-state combination differences of CH$_5^+$}.
\newblock {\em Science}, 347:1346--1349, 2015.

\bibitem{15WoMa}
Robert Wodraszka and Uwe Manthe.
\newblock Ch$_5^+$: Symmetry and the entangled rovibrational quantum states of
  a fluxional molecule.
\newblock {\em J. Phys. Chem. Lett.}, 6(21):4229--4232, 2015.

\bibitem{15ScScJe}
Hanno Schmiedt, Stephan Schlemmer, and Per Jensen.
\newblock Symmetry of extremely floppy molecules: Molecular states beyond
  rotation-vibration separation.
\newblock {\em J. Chem. Phys.}, 143(15):154302, 2015.

\bibitem{16WaCa}
Xiao-Gang Wang and Tucker Carrington.
\newblock Calculated rotation-bending energy levels of ch$_5^+$ and a
  comparison with experiment.
\newblock {\em J. Chem. Phys.}, 144(20):204304, 2016.

\bibitem{17BrScAs}
Stefan Brackertz, Stephan Schlemmer, and Oskar Asvany.
\newblock Searching for new symmetry species of ch$_5^+$ -- from lines to
  states without a model.
\newblock {\em J. Mol. Spectrosc.}, 342(Supplement C):73 -- 82, 2017.

\bibitem{17FaQuCs}
C.~F\'abri, M.~Quack, and A.~G. Cs\'asz\'ar.
\newblock {On the use of nonrigid-molecular symmetry in nuclear-motion
  computations employing a discrete variable representation: a case study of
  the bending energy levels of CH$_5^+$}.
\newblock {\em J. Chem. Phys.}, 147:134101, 2017.

\bibitem{18FaCs}
C.~F\'abri and A.~G. Cs\'asz\'ar.
\newblock {Vibrational quantum graphs and their application to the quantum
  dynamics of CH$_5^+$}.
\newblock {\em Phys. Chem. Chem. Phys.}, 20:16913--16917, 2018.

\bibitem{20CsFaSa}
Attila~G. Cs\'asz\'ar, Csaba F\'abri, and J\'anos Sarka.
\newblock Quasistructural molecules.
\newblock {\em WIREs Comput. Mol. Sci.}, 10(1):e1432, 2020.

\bibitem{84NaEzDeBe}
Grigory~A. Natanson, Gregory~S. Ezra, Gerardo Delgado-Barrio, and R.~Stephen
  Berry.
\newblock Calculation of rovibrational spectra of water by means of
  particles‐on‐concentric‐spheres models. i. ground stretching
  vibrational state.
\newblock {\em J. Chem. Phys.}, 81(8):3400--3406, 1984.

\bibitem{86NaEzDeBe}
Grigory~A. Natanson, Gregory~S. Ezra, Gerardo Delgado-Barrio, and R.~Stephen
  Berry.
\newblock Calculation of rovibrational spectra of water by means of
  particles‐on‐concentric‐spheres models. ii. excited states of
  stretching vibrations.
\newblock {\em J. Chem. Phys.}, 84(4):2035--2044, 1986.

\bibitem{88LeNaBeVi}
David~M. Leitner, Grigory~A. Natanson, R.Stephen Berry, Pablo Villarreal, and
  Gerardo Delgado-Barrio.
\newblock Particles-on-a-sphere method for computing the rotational-vibrational
  spectrum of h$_2$o.
\newblock {\em Comput. Phys. Commun.}, 51(1):207 -- 216, 1988.

\bibitem{88HuLeNaBe}
John~E. Hunter, David~M. Leitner, Grigory~A. Natanson, and R.Stephen Berry.
\newblock Theoretical intensities for rotation-vibration lines of water using
  particles-on-a-sphere wavefunctions.
\newblock {\em Chem. Phys. Lett.}, 144(2):145 -- 148, 1988.

\bibitem{05DeNe}
Michael~P. Deskevich and David~J. Nesbitt.
\newblock Large amplitude quantum mechanics in polyatomic hydrides. i. a
  particles-on-a-sphere model for xh$_n$.
\newblock {\em J. Chem. Phys.}, 123(8):084304, 2005.

\bibitem{08DeMcHuNe}
Michael~P. Deskevich, Anne~B. McCoy, Jeremy~M. Hutson, and David~J. Nesbitt.
\newblock {Large-amplitude quantum mechanics in polyatomic hydrides. II. A
  particle-on-a-sphere model for XH$_n$ ($n=4,5$)}.
\newblock {\em J. Chem. Phys.}, 128:094306, 2008.

\bibitem{14UhWaFo}
Felix Uhl, \L{}ukasz Walewski, Harald Forbert, and Dominik Marx.
\newblock Adding flexibility to the “particles-on-a-sphere” model for
  large-amplitude motion: Posflex force field for protonated methane.
\newblock {\em J. Chem. Phys.}, 141(10):104110, 2014.

\bibitem{16ScJeSc}
H.~Schmiedt, P.~Jensen, and S.~Schlemmer.
\newblock Collective molecular superrotation: A model for extremely flexible
  molecules applied to protonated methane.
\newblock {\em Phys. Rev. Lett.}, 117:223002, 2016.

\bibitem{17ScJeSc}
Hanno Schmiedt, Per Jensen, and Stephan Schlemmer.
\newblock Rotation-vibration motion of extremely flexible molecules -- the
  molecular superrotor.
\newblock {\em Chem. Phys. Lett.}, 672(Supplement C):34 -- 46, 2017.

\bibitem{17ScJeSc2}
Hanno Schmiedt, Per Jensen, and Stephan Schlemmer.
\newblock The role of angular momentum in the superrotor theory for
  rovibrational motion of extremely flexible molecules.
\newblock {\em J. Mol. Spectrosc.}, 342(Supplement C):132 -- 137, 2017.

\bibitem{19Rawlinso}
J.~I. Rawlinson.
\newblock Quantum graph model for rovibrational states of protonated methane.
\newblock {\em J. Chem. Phys.}, 151(16):164303, 2019.

\bibitem{36Pauling}
Linus Pauling.
\newblock The diamagnetic anisotropy of aromatic molecules.
\newblock {\em J. Chem. Phys.}, 4(10):673--677, 1936.

\bibitem{18Rawlinso}
J.~I. Rawlinson.
\newblock An alpha particle model for carbon-12.
\newblock {\em Nucl. Phys. A}, 975:122 -- 135, 2018.

\bibitem{20Rawlinso}
J.~I. Rawlinson.
\newblock {\em Rovibrational Dynamics of Nuclei and Molecules}.
\newblock PhD thesis, University of Cambridge, 2020.

\bibitem{13BeKu}
Gregory Berkolaiko and Peter Kuchment.
\newblock {\em Introduction to Quantum Graphs}, volume 186.
\newblock American Mathematical Society, 2013.

\bibitem{06BuJe}
P.~R. Bunker and P.~Jensen.
\newblock {\em {Molecular symmetry and spectroscopy}}.
\newblock NRC Research Press, Ottawa, 2006.

\bibitem{93ScKiSc}
Peter~R. Schreiner, Seung-Joon Kim, Henry~F. Schaefer~III, and Paul von
  Ragu\'e~Schleyer.
\newblock Ch$_5^+$: The never‐ending story or the final word?
\newblock {\em J. Chem. Phys.}, 99(5):3716--3720, 1993.

\bibitem{13WiIvMa}
Alexander Witt, Sergei~D. Ivanov, and Dominik Marx.
\newblock {Microsolvation-induced quantum localization in protonated methane}.
\newblock {\em Phys. Rev. Lett.}, 110:083003, 2013.

\bibitem{93Bi}
N.~Biggs.
\newblock {\em Algebraic Graph Theory}.
\newblock Cambridge University Press, 1993.

\bibitem{16ArFuCs}
P.~\'Arend\'as, T.~Furtenbacher, and A.~G. Cs\'asz\'ar.
\newblock {On spectra of spectra}.
\newblock {\em J. Math. Chem.}, 54:806--822, 2016.

\end{thebibliography}

\end{document}